\documentclass[aps,prd,nofootinbib,onecolumn,superscriptaddress,preprintnumbers,balancelastpage,longbibliography,nobibnotes]{revtex4-1}
\usepackage{graphicx,epsfig,psfrag,bm,amssymb}
\usepackage{dcolumn}
\usepackage{bm}
\usepackage[dvipsnames]{xcolor}
\usepackage{braket}
\usepackage{mathrsfs,amsfonts,hepunits,color}
\usepackage{hyperref}
\usepackage{comment}
\usepackage{soul}

\usepackage{subcaption}
\usepackage{tikz-feynman}
\usepackage{tikz}
\usetikzlibrary{decorations.pathmorphing}
\usetikzlibrary{decorations.pathmorphing} 

\newcommand{\be}{\begin{equation}}
\newcommand{\ee}{\end{equation}}
\newcommand{\bea}{\begin{eqnarray}}
\newcommand{\eea}{\end{eqnarray}}

\definecolor{linkcolor}{rgb}{0.0, 0.28, 0.67}
\hypersetup{colorlinks=true,
	linkcolor=black,
	citecolor=linkcolor,
	urlcolor=linkcolor,
	linktocpage=true
}

\DeclareRobustCommand{\Eq}[1]{Eq.~(\ref{eq:#1})}
\DeclareRobustCommand{\Eqs}[2]{Eqs.~(\ref{eq:#1}) and (\ref{eq:#2})}
\newcommand{\Sec}[1]{Sec.~\ref{sec:#1}}
\newcommand{\App}[1]{Appendix~\ref{app:#1}}
\newcommand{\Fig}[1]{Fig.~\ref{fig:#1}}

\newcommand{\h}{h}

\newcommand{\Tmax}{T_{\rm max}}
\newcommand{\Mp}{m_{\rm p}}

\newcommand{\LEH}{\mathcal{L}_{\rm EH}}
\newcommand{\Lphi}{\mathcal{L}_{\phi }}

\newcommand{\fh}{f_\h}
\newcommand{\fp}{f_{\phi}}
\newcommand{\dfh}{\dot{f}_\h}
\newcommand{\dfp}{\dot{f}_{\phi}}

\newcommand{\gss}{g_{*\text{s}}}
\newcommand{\gsr}{g_{*\rho}}
\newcommand{\gsc}{g_{*\text{c}}}
\newcommand{\gs}{g_{*}}

\newcommand{\fg}{f_{\rm g}}

\newcommand{\ymax}{y_{\rm max}}

\newcommand{\OmGW}{\Omega_{\rm gw}}
\newcommand{\rhoGW}{\rho_{\rm gw}}

\newcommand{\Lh}{L_\h}
\newcommand{\Gh}{G_\h}
\newcommand{\Lp}{L_{\phi}}
\newcommand{\Gp}{G_{\phi}}

\newcommand{\vp}{\mathbf{p}}
\newcommand{\vk}{\mathbf{k}}

\newcommand{\Ott}{\mathcal{O}\left(\lambda^2\kappa^2\right)}
\newcommand{\Ozf}{\mathcal{O}\left(\kappa^4\right)}
\newcommand{\Otf}{\mathcal{O}\left(\lambda^2\kappa^4\right)}

\newcommand{\Ooo}{\mathcal{O}\left(\lambda^1\kappa^1\right)}
\newcommand{\Ooth}{\mathcal{O}\left(\lambda^1\kappa^3\right)}

\newcommand{\Tphi}{T_{\phi}}

\newcommand{\rmii}[1]{{\mbox{\tiny\rm{#1}}}}

\newcommand{\nB}{n_\rmii{B}}
\def\nb{\nB}

\begin{document}
	
	\title{Freezing-In Gravitational Waves}
	\author{Jacopo Ghiglieri}
	\affiliation{SUBATECH, Universit\'e de Nantes, IMT Atlantique, IN2P3/CNRS\\
 4 rue Alfred Kastler, La Chantrerie BP 20722, 44307 Nantes, France}
	\author{Jan Sch\"{u}tte-Engel}
	\author{Enrico Speranza}
	\affiliation{Department of Physics, University of Illinois at Urbana-Champaign, Urbana, IL 61801, USA}
	\affiliation{Illinois Center for Advanced Studies of the Universe, University of Illinois at Urbana-Champaign, Urbana, IL 61801, USA}

	\begin{abstract}
		The thermal plasma in the early universe produced a stochastic gravitational wave (GW) background, which peaks today in the microwave regime and was dubbed the cosmic gravitational microwave background (CGMB).
		In previous works only single graviton production processes that contribute to the CGMB have been considered. 
		Here we also investigate graviton pair production processes and show that these can lead to a significant contribution if the ratio between the maximum temperature and the Planck mass, $T_{\rm max}/m_{\rm p}$, divided by the internal coupling in the heat bath is large enough.
        As the dark matter freeze-in production mechanism is conceptually very similar to
	    the GW production mechanism from the primordial thermal plasma, we refer to the latter as ``GW freeze-in production''.
        We show that quantum gravity effects appear in single graviton production and are smaller by a factor $(T_{\rm max}/m_{\rm p})^2$ than the leading order contribution.
        In our work we explicitly compute the CGMB spectrum within a scalar model with quartic interaction.
    
	\end{abstract}

	\maketitle
	
\section{Introduction}
The first detection of gravitational waves (GWs) from back hole and neutron star mergers~\cite{PhysRevLett.116.061102,PhysRevLett.119.161101} opened up a new window to explore our universe. 
While the GWs that have been detected so far were emitted in the late-time universe, GWs can also be produced in the early universe. These GWs are stochastic in nature and their detection would yield unprecedented information about early universe cosmology as well as high energy particle physics. To give a few examples, GWs in the early universe can be produced from inflation~\cite{Grishchuk:1975abc,Starobinskii:1979abc,RUBAKOV1982189,1983PhLB..125..445F}, preheating~\cite{Khlebnikov:1997di,Lozanov}, inflaton annihilation into gravitons~\cite{Ema:2015dka,Ema:2016hlw,Ema:2020ggo}, 
first-order phase transitions~\cite{Witten:1984rs,Hogan:1986qda}, cosmic defects such as cosmic strings~\cite{Damour:2000wa,Damour:2001bk}, noisy turbulent motion~\cite{Kosowsky:2001xp,Nicolis:2003tg,Caprini:2006jb,Gogoberidze:2007an,Kalaydzhyan:2014wca} and equilibrated gravitons~\cite{Kolb:206230,Vagnozzi:2022qmc}.
For a review on early universe GW sources see Ref.~\cite{Caprini:2018mtu}. The full GW spectrum for a specific particle physics model that can describe the entire cosmological history was worked out in Ref.~\cite{Ringwald:2022xif}.

In this paper we consider GWs that were produced from the thermal plasma~\cite{Weinberg:1972kfs,Ghiglieri:2015nfa,Ghiglieri:2020mhm,Ringwald:2020ist} in the early universe. 
Every plasma, even in thermal equilibrium, produces GWs due to microscopic particle collisions and macroscopic hydrodynamic fluctuations, cf. Ref.~\cite{Ghiglieri:2015nfa}. In the former case the GW momenta are on the order
of the temperature, $k\sim T$ and in the latter case they are much smaller, $k\ll T$.  Here, we focus on GWs produced by microscopic particle collisions, since it enables us to probe  elementary particle physics theories at high energies.  Furthermore, the GW contribution from microscopic particle collisions to the final spectrum is larger compared to the contribution from hydrodynamic fluctuations~\cite{Ghiglieri:2015nfa}. 

Our main assumption is that after the hot Big Bang a thermal plasma of particles in thermal equilibrium at a maximum temperature $\Tmax$ was present. In addition we assume that at this time no GWs are present. 
In an expanding universe GWs from the thermal plasma are continuously produced as the temperature decreases.
The spectrum of the produced GWs peaks at a frequency on the order of the temperature at the time of production. If the redshift-temperature relation is linear, 
the GW spectra that are produced at different temperatures add up such that the observed GW spectrum today is enhanced.
The spectrum of the produced GWs today peaks in the microwave regime and is hence dubbed the cosmic gravitational microwave background (CGMB).

In principle, the maximum temperature, $\Tmax$, of the thermal plasma can be as high as the Planck mass $\Mp\approx1.2\times 10^{19}\,$ GeV~\cite{Sakharov:1966fva}. However in slow roll inflationary cosmology it cannot be much higher than $10^{-3}\,\Mp$. This bound follows by first inferring the energy scale of inflation from the amplitude of scalar perturbations and the tensor-to-scalar ratio and then assuming an instantaneous and a maximally efficient reheating~\cite{Felder:1998vq} to a radiation dominated universe, cf. Ref.~\cite{Ringwald:2020ist} and Ref. \cite{Amin:2014eta} for a review.
Note that $\Tmax>10^{-3}\,\Mp$ can be achieved in non-inflationary scenarios. One particular example are bouncing cosmology scenarios which can lead to $\Tmax$ which goes up to the Planck scale: $\Tmax< \Mp$, cf. Refs.~\cite{Brandenberger:2016vhg,Brandenberger:2020tcr,Hu:2020wul}.
The maximum temperature of the thermal plasma is also bounded from below. The most conservative estimates 
set a lower limit around a few MeV~\cite{Kawasaki:1999na,Kawasaki:2000en,Giudice:2000ex,Hannestad:2004px,Hasegawa:2019jsa}, shortly before
Big Bang nucleosynthesis  took place. However, most scenarios require 
temperatures reaching well above the electroweak scale such that, e.g., sphalerons can be active in leptogenesis scenarios~\cite{Fukugita:1986hr}.

In previous works the CGMB spectrum has been calculated within the Standard Model (SM)~\cite{Ghiglieri:2015nfa,Ghiglieri:2020mhm} and for Beyond Standard Model (BSM) theories~\cite{Ringwald:2020ist,Castells-Tiestos:2022qgu,Klose:2022knn,Klose:2022rxh}. 
Those works considered only \textit{single graviton production} processes. 
In this case the resulting GW energy density per logarithmic momentum interval, $\OmGW$, is  proportional to  $g^2\times(\Tmax/\Mp)$, where $g$ is the internal coupling in the thermal bath.
Here, we extend previous works by also including GW production processes with two gravitons in the final state. These give a contribution to $\OmGW$ that is proportional to $\left(\Tmax/\Mp\right)^3$. 
We refer to this GW production channel as \textit{graviton pair production}.
Depending on the values of $\Tmax/\Mp$ and $g$, the graviton pair production channel can be the dominating contribution to the CGMB spectrum.
In analogy to dark matter production from the thermal plasma we dub the GW production from the thermal plasma \textit{GW freeze-in production}.

We also identify at which order
quantum gravity and back-reaction effects would appear in the CGMB spectrum. Observing these effects in the CGMB spectrum would therefore probe the quantization of gravity and reveal fundamental information about particle physics, if the GW production occurs at high energy scales that cannot be probed with particle colliders on Earth.

Throughout this paper we work with a complex scalar field with quartic coupling $\lambda$ that is the internal coupling in the thermal bath, i.e., for our model the previously mentioned generic coupling $g$ is the quartic coupling $\lambda$.
In previous works~\cite{Ghiglieri:2015nfa,Ghiglieri:2020mhm,Ringwald:2020ist} such a coupling has not been considered even though the SM has such a coupling in the Higgs sector. 
That is because Refs.~\cite{Ghiglieri:2015nfa,Ghiglieri:2020mhm}
worked under the assumption that in the SM the three gauge couplings and the top Yukawa are of order of the
square root of the Higgs self-coupling. Note that in BSM theories this is not necessarily the case.

This paper is organized as follows: in \Sec{scalar_model} we introduce our model, which is a complex scalar field coupled to gravity. This is then followed by \Sec{evolution_eqs}, where we introduce the full evolution equations for the two distribution functions $\fp$ and $\fh$ which describe the scalars and gravitons, respectively. Furthermore,
we perturbatively expand the distribution functions around their initial states. This enables us to find a solution of the coupled nonlinear integral-differential equations for the distribution functions. In \Sec{Matrix_elements} we calculate the Matrix elements squared for the graviton production processes. We then compute the GW spectrum in  \Sec{embeeding_in_cosmology}. Finally, conclusions are given in  \Sec{conclusion}. Throughout this paper we use natural units with $\hbar=c=k_{\rm B}=1$, where $k_{\rm B}$ is the Boltzmann constant.

\section{Scalar model}\label{sec:scalar_model}
The action for a complex scalar field on curved space-time is
\begin{eqnarray}
S_{\phi}=\int d^4x\ \Lphi=\int d^4x\, \sqrt{-g}\left(-g^{\mu\nu}\left(\nabla_\mu\phi\right)^\dagger\nabla_\nu\phi-U\right),
\label{eq:Model_action}
\end{eqnarray}
where $\nabla_\mu$ is a covariant derivative, $g_{\mu\nu}$ the metric tensor, $g={\rm Det}\left[g_{\mu\nu}\right]$ and $U$ is the potential.
The flat space-time metric is defined as $\eta_{\mu\nu}\equiv{\rm diag}(-1,1,1,1)$. 
Note that for a scalar field the covariant derivative reduces to a partial derivative: $\nabla_\mu\phi=\partial_\mu \phi$. The considered complex scalar is not charged under a local transformation and we consider a quartic potential $U=-\frac{\lambda}{4}|\phi|^4$. We assume that the scalar field is massless, which is justified if the considered temperature in the thermal plasma is larger than the mass of the scalar field. 

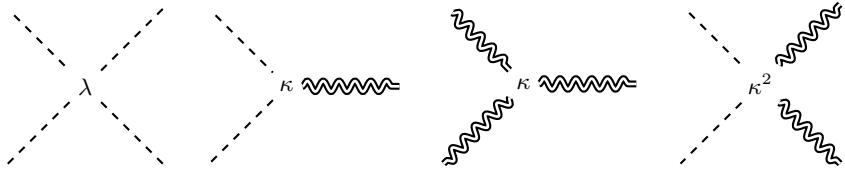
\begin{figure}[t]
	
	\begin{tikzpicture}
	\begin{feynman}
	\vertex (a) {};
	\vertex [below right=of a] (b){\(\lambda\)};
	\vertex [above right=of b] (c);
	\vertex [below left=of b] (d) ;	
	\vertex [below right=of b] (e);
	
	\diagram* {
		(a) -- [scalar,thick] (b) -- [scalar,thick] (c),
		(b) -- [scalar,thick] (d), (b) -- [scalar,thick] (e)
	};
	\end{feynman}
	\end{tikzpicture}
	~~~\begin{tikzpicture}
	\begin{feynman}
	\vertex (a) {};
	\vertex [below right=of a] (b) {\(\kappa\)};
	\vertex [right=of b] (c);
	\vertex [below left=of b] (d) ;

	\diagram* {
		(a) -- [scalar,thick] (b) -- [thick, double distance=1pt,decoration={snake, amplitude=2pt, segment length=6pt}, decorate] (c),
		(b) -- [scalar,thick] (d)
	};
	\end{feynman}
	\end{tikzpicture}
	~~~\begin{tikzpicture}
	\begin{feynman}
	\vertex (a) {};
	\vertex [below right=of a] (b) {\(\kappa\)};
	\vertex [right=of b] (c);
	\vertex [below left=of b] (d) ;

	\diagram* {
		(a) -- [thick, double distance=1pt,decoration={snake, amplitude=2pt, segment length=6pt}, decorate] (b) -- [thick, double distance=1pt,decoration={snake, amplitude=2pt, segment length=6pt}, decorate] (c),
		(b) -- [thick, double distance=1pt,decoration={snake, amplitude=2pt, segment length=6pt}, decorate] (d)
	};
	\end{feynman}
	\end{tikzpicture}
	~~~\begin{tikzpicture}
	\begin{feynman}
	\vertex (a) {};
	\vertex [below right=of a] (b) {\(\kappa^2\)};
	\vertex [above right=of b] (c);
	\vertex [below right=of b] (d) ;
	\vertex [below left=of b] (e) ;

	\diagram* {
		(a) -- [scalar,thick] (b) --  [thick, double distance=1pt,decoration={snake, amplitude=2pt, segment length=6pt}, decorate] (c),
		(b) -- [thick, double distance=1pt,decoration={snake, amplitude=2pt, segment length=6pt}, decorate] (d), (b) -- [scalar,thick] (e)
	};
	\end{feynman}
	\end{tikzpicture}
	\caption{Lowest order vertices that arise from the expansion of the scalar field and Einstein-Hilbert Lagrangian. Scalars are represented by dashed lines and gravitons by double lines.}
		\label{fig:vertices}
\end{figure}

On top of the action in \Eq{Model_action} we need the Einstein-Hilbert action
\begin{eqnarray}
S_{\rm EH}=\int d^4x\, \mathcal{L}_{\rm EH}=\int d^4x\,\frac{1}{16\pi G} \sqrt{-g}\,R,
\end{eqnarray}
where $R$ is the Ricci scalar and $G\equiv 1/\Mp^2$ is the gravitational constant.
In the following we expand the metric around flat-space-time: $g_{\mu\nu}=\eta_{\mu\nu}+h_{\mu\nu}$, with $h_{\mu\nu}\ll 1$. A detailed expansion has been worked out before in Ref.~\cite{Choi:1994ax} and yields
\begin{eqnarray}	
\mathcal{L}=\LEH+\Lphi&=&
-\eta^{\mu\nu}\left(\partial_\mu\phi\right)^\dagger\partial_\nu\phi-U
+\frac{1}{2}\partial_\mu h^{\sigma\nu}\partial^\mu h_{\sigma\nu}\nonumber\\
&~&+\kappa\, h^{\mu\nu}\left(\partial_\mu\phi\right)^\dagger\partial_\nu\phi
+\kappa\left(-\frac{1}{2}h^{\alpha}_{~\beta}\partial_\alpha h^{\mu}_{~\nu}\partial^\beta h^\nu_{~\mu}
-h^{\alpha}_{~\beta}\partial_\mu h^{\nu}_{~\alpha}\partial^\mu h^\beta_{~\nu}
+h^\beta_{~\mu}\partial_\nu h^\alpha_{~\beta}\partial^\mu h^{\nu}_{~\alpha}\right)\nonumber\\
&~&+\kappa^2\left[\left(-h^{\mu\lambda}h_{\lambda}^{~\nu}+\frac{1}{4}\eta^{\mu\nu}h^{\alpha}_{~\rho}h^\rho_{~\alpha}\right)\left(\partial_\mu\phi\right)^\dagger\partial_\nu\phi+\frac{1}{4}h^{\alpha}_{~\rho}h^\rho_{~\alpha}\,U\right]
+\mathcal{O}(h^3).	
\label{eq:Lagrangian_linearized_gravits}
\end{eqnarray}
Note that the $h$-fields in~\Eq{Lagrangian_linearized_gravits} have been rescaled with a factor $\kappa\equiv\sqrt{32\pi G}$ and have now mass dimension one.
Furthermore, we have adopted the so-called transverse-traceless (TT) gauge, which includes the De Donder gauge: $\partial_\alpha h^{\alpha}_{~\mu}=\frac{1}{2}\partial_\mu h$ together with the requirement that the trace $h=h^\mu_{~\mu}$ is zero. 
In the first and second line of \Eq{Lagrangian_linearized_gravits} we wrote down the zeroth and first order terms coming from $\LEH$ and $\Lphi$. In the second line we only write down the second order term coming from $\Lphi$, since the second order term from $\LEH$ will not be needed for our calculations.
The lowest order Feynman vertices for our theory are shown in \Fig{vertices}.

\section{Evolution equations for the distribution functions}\label{sec:evolution_eqs}

We describe the thermal plasma of $\phi$-particles and the produced gravitons with two distribution functions defined as
\begin{eqnarray}
\label{fdistr_1}
\fp(t,k)\equiv\frac{N_{\phi}^k}{V\, d^3k/(2\pi)^3}\label{eq:f_h_definition},~~~~
\fh(t,k)\equiv\frac{N_\h^k}{V \,d^3k/(2\pi)^3},
\end{eqnarray}
$V$ is the considered volume, and $N_{\phi}^k$ and $N_\h^k$  
are the numbers of $\phi$-states and gravitons  with momentum $k=|\vk|$ in the interval $d^3k$.
Since we shall expand $\fp$ around an isotropic 
equilibrium state, $\fh$ is understood to be the polarization-averaged distribution function. Also we do not introduce a distribution function for $\phi^\dagger$ since our model and initial conditions are CP symmetric and therefore it would always be equal to $\fp$.

In the regime where the momentum $k$ is
on the order of the hard scale which in equilibrium corresponds
to the temperature, i.e. $k\sim T$, kinetic theory is expected to be a good approximation for our system. 
The evolution equations for the $\phi$ and graviton distribution functions can thus be written in the following Boltzmann-like form
\begin{eqnarray}
\dfp(t,k)&=&\Gp(t,k)-\Lp(t,k),\label{eq:evolution_eq_fphi} \\
\dfh(t,k)&=&\Gh(t,k)-\Lh(t,k),\label{eq:evolution_eq_fh}
\end{eqnarray}
where the
$G$ and $L$ terms describe the gain and loss terms of particle states. 
A generic expression for the graviton production term $\Gh$ is given by
\begin{eqnarray}
\Gh(t,k)=\frac{1}{4k}\sum_{\substack{\text{all processes }r\\ \text{with at least one }\\\text{final state graviton}}}S_{r}&\,&\int d\Omega_{r} \, \left|\mathcal{M}_{r}\right|^2 \times \fp(p_1')\cdots \fp(p_m')\,\fh(k_1')\cdots \fh(k_n')\times \nonumber\\
&~& \times\left(1+\fp(p_1)\right)\cdots\left(1+\fp(p_i)\right)\, \left(1+\fh(k)\right)\cdots \left(1+\fh(k_j)\right),
\label{eq:Gh}
\end{eqnarray}
where the index $r$ labels all possible processes.  
We call the momenta of the incoming $\phi$ and graviton states $p_1',\cdots, p_m'$ and $k_1',\cdots, k_n'$, respectively. The momenta of the outgoing $\phi$'s and gravitons are $p_1,\cdots, p_i$ and $k_1=k,k_2,\cdots ,k_j$. In our notation the $m$ incoming and $i$ outgoing $\phi$-states can be $\phi$ or $\phi^\dagger$ states. In \Eq{Gh} the symmetry factor $S_{r}$ has to be included if two or more indistinguishable particles appear in the initial or final state. We will make the symmetry factor explicit in section \Sec{Matrix_elements} where we calculate the graviton rate.
The sum in \Eq{Gh} runs over combinations of all processes with at least one graviton with momentum $k$ in the final state. The pre-factor ${1}/({4k)}$ is a combination of ${1}/{(2k)}$ from the phase space measure and  ${1}/{2}$ from the graviton polarization degeneracy. We need the factor $2$ from the polarization degeneracy since the matrix element squared is summed over  polarizations and the distribution function is defined to be averaged over both polarizations. 
The loss term is analogous to the gain term in \Eq{Gh} with the difference that one sums over all processes with at least one graviton in the initial state.
The Boltzmann-like Eqs. \eqref{eq:evolution_eq_fphi}, \eqref{eq:evolution_eq_fh} and \eqref{eq:Gh} come with important caveats on their validity beyond leading order, which we shall discuss later.
The integral that appears in $\Gh$ is the phase space integral that has to be performed over all momenta, except $k$:
\begin{eqnarray}
\int d\Omega_{r}&=&\int\frac{d^3 p_1'}{(2\pi)^3 2p_1'}\cdots\int\frac{d^3 p_m'}{(2\pi)^3 2p_m'}\, \int\frac{d^3 k_1'}{(2\pi)^3 2k_1'}\cdots\int\frac{d^3 k_n'}{(2\pi)^3 2k_n'}\, \int\frac{d^3 p_1}{(2\pi)^3 2p_1}\cdots\int\frac{d^3 p_i}{(2\pi)^3 2p_i}\, \, \times  \nonumber\\
&~&\times \int\frac{d^3 k_2}{(2\pi)^3 2k_2}\cdots\int\frac{d^3 k_j}{(2\pi)^3 2k_j}\,\times\,(2\pi)^4\, \delta^{(4)}(P_1'\cdots+P_m'+K_1'\cdots + K_n'-P_1\cdots -P_i-K\cdots-K_j),
\label{eq:phasespace}
\end{eqnarray}
where we use capital letters to denote four-vectors. For further use we introduce the shorthand notation: 
\begin{eqnarray}
\int_{r}:=\frac{S_{r}}{4k}\int d\Omega_{r}\, \left|\mathcal{M}_{r}\right|^2.
\end{eqnarray}
We have written Eqs.~\eqref{eq:evolution_eq_fphi}, \eqref{eq:evolution_eq_fh}, \eqref{eq:Gh} and~\eqref{eq:phasespace} in a rather generic form which includes all possible processes. 
In our specific model of a massless complex scalar field,
$1\leftrightarrow 2$ processes
are only allowed in the collinear limit, i.e.,
when the three-momenta of all three particles 
are exactly parallel.
However, 
in this case the thermal and vacuum masses of 
the scalars have to be taken into account, which   leads to the fact that
$1\leftrightarrow 2$ graviton production processes are not even allowed in the collinear limit. The first kinematically
allowed processes for graviton production are  $2\to 2$ and $2\leftrightarrow 3$ processes, which have a lowest order matrix element squared of $\Ozf$ and $\Ott$, respectively.

We are interested in tracking the evolution of $\fh$ under the assumption that it starts from an initially vanishing value 
in a bath of equilibrated scalars.\footnote{
Although we consider the case of an initial vanishing distribution function of gravitons, our framework
is more generic, as it is valid as long as $f_h\ll 1$.} 
In our stated freeze-in scenario, we further assume 
that throughout the entire evolution $\fh\ll 1$ and $|\nb-\fp|\ll1$, where 
$\nb$ is the Bose--Einstein distribution $\nb(k)\equiv 1/(e^{k/T}-1)$.
We can thus expand the distribution functions up to fourth order in $\lambda$ and $\kappa$:
\begin{eqnarray}
\fp(k)&=&\nb(k)+\fp^{(2,2)}(k),\label{eq:fphi_expansion}\\
\fh(k)&=&0+\fh^{(2,2)}(k)+\fh^{(0,4)}(k),\label{eq:fh_expansion}
\end{eqnarray}
where 
the superscript stands for the order of $\lambda$ and $\kappa$ that is considered, i.e. $\mathcal{O}(\lambda^i\kappa^j)=(i,j)$.
Note that in our expansion of the distribution functions we have also implicitly expanded the matrix element squared.
As $\kappa$ is dimensionful, the expansion in $\kappa\sim1/\Mp$
has to be understood as an expansion in $T\kappa$, the corresponding dimensionless quantity. 
The zeroth order term of $\fp$ is set to be $\nb$ and the zeroth order term of the $\fh$ distribution function is zero. We have suppressed the time arguments in \Eqs{fphi_expansion}{fh_expansion}.
Note that in the expansion we treat $\lambda$ and $T\kappa$ on equal footing and for the distribution functions we have only written out the non-vanishing terms up to fourth order.
Terms of $\mathcal{O}\left(\lambda^{n}\kappa^{2}\right)$ with $n>2$  are also
expected, which  --- depending on the temperature and the value for $\lambda$ ---  can be larger than $\mathcal{O}\left(\kappa^4\right)$. However, as we 
discuss later in this section, these terms cannot in general be included in a straightforward manner into the  Boltzmann-like ansatz, cf. Eqs.~\eqref{eq:evolution_eq_fphi},~\eqref{eq:evolution_eq_fh} and \eqref{eq:Gh}.

In the following, we discuss based on three examples why the terms shown in \Eqs{fphi_expansion}{fh_expansion} are the only non-zero contributions.
The evolution Eqs. \eqref{eq:evolution_eq_fphi} and \eqref{eq:evolution_eq_fh}  yield $\dfh^{(2,0)}=0$ since, in order to produce or annihilate a graviton, one has to go at least to second order in $\kappa$. Furthermore,  $\dot{f}_h^{(0,2)}=0=\dot{f}_{\phi}^{(0,2)}$, since massless
$1\leftrightarrow 2$ processes are kinematically forbidden.
Finally, $\dot{f}_\phi^{(2,0)}$ vanishes because of 
detailed balance arguments. As an example consider the two terms $\int_{\phi \phi\to\phi\phi}^{(2,0)}\nb\nb(1+\nb)(1+\nb(k))-\int_{\phi\phi\to\phi\phi}^{(2,0)}\nb(k) \nb (1+\nb)(1+\nb)$ that appear in $\dfp^{(2,0)}$. After a redefinition of variables in the phase space integral one can show that both terms cancel each other. Similar arguments hold for the other terms in $\dfp^{(2,0)}$ such that overall $\dfp^{(2,0)}=0$. 
Note that if $\dot{f}^{(i,j)}=0$ then $f^{(i,j)}=0$ for all times for $(i,j)\neq(0,0)$ which follows from the initial conditions: $\fh(t=0)=0$ and $\fp(t=0)=\nb$.
In complete analogy to the discussed examples one can show with kinematic and detailed balance arguments that the other terms that are not shown in \Eqs{fphi_expansion}{fh_expansion} also vanish.

\begin{figure}[t]
	\centering
	\begin{subfigure}[c]{.25\textwidth}
	\begin{tikzpicture}
	\begin{feynman}
	\vertex (a) {\(\phi \)};
	\vertex [below right=1.35cm of a] (b) ;
	\vertex [above right=1.1cm of b] (c){\(h \)};
	\vertex [below=1.1cm of b] (d) ;
	\vertex [above right=1.1cm of d] (e) {\(\phi \)};
	\vertex [below right=1.1cm of d] (f) {\(\phi \)};
	\vertex [below left=1.1cm of d] (g)  {\(\phi \)};	
	\diagram* {
		(b) -- [scalar,thick] (a), 
		(g) -- [scalar,thick] (d),
		(b) -- [thick, double distance=1pt,decoration={snake, amplitude=2pt, segment length=6pt}, decorate] (c),
		(b) -- [scalar,thick] (d),
		(d) -- [scalar,thick] (e),
		(d) -- [scalar,thick] (f)
	};
	~~~~~~\end{feynman}
	\end{tikzpicture}
	\end{subfigure}
	\begin{subfigure}[c]{.25\textwidth}
	\begin{tikzpicture}
	\begin{feynman}
	\vertex (a) {\(\phi \)};
	\vertex [below right=1.35cm of a] (b) ;
	\vertex [right=1.1cm of b] (c);
	\vertex [above right=1.1cm of c] (d) {\(h \)} ;
	\vertex [right=1.1cm of c] (e) {\(\phi \)};
	\vertex [below right=1.1cm of b] (f) {\(\phi \)};
	\vertex [below left=1.1cm of b] (g)  {\(\phi \)};	
	\diagram* {
		(b) -- [scalar,thick] (a), 
		(b) -- [scalar,thick] (c),
		(c) -- [thick, double distance=1pt,decoration={snake, amplitude=2pt, segment length=6pt}, decorate] (d),
		(c) -- [scalar,thick] (e),
		(b) -- [scalar,thick] (f),
		(b) -- [scalar,thick] (g)
	};
	\end{feynman}
	\end{tikzpicture}
	\end{subfigure}
	
	\caption{Lowest order Feynman diagrams for the $\phi\phi \to\phi \phi h$ process. The matrix element squared is of the order $\mathcal{O}\left(\lambda^2\kappa^2\right)$. In addition to the diagrams shown, there are two more diagrams that can be obtained by crossing the final state graviton.}
	\label{fig:Ph}
\end{figure}
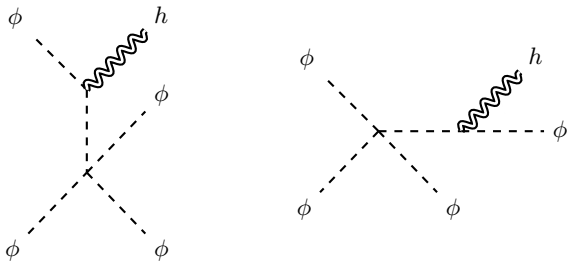

Next we discuss the non-zero fourth order terms for the graviton distribution function. 
The $\dot{f}_h^{(2,2)}$ rate is non-zero and it is given by
\begin{eqnarray}
	\dfh^{(2,2)}= \int_{\phi\phi \to\phi \phi h}^{(2,2)} \nb \nb (1+\nb) (1+\nb)+\cdots ,
	\label{eq:dfh22}
\end{eqnarray}
where the dots stand for other processes of the same order, e.g.,  $\phi\phi^\dagger\phi\to\phi h$. All possible processes are written down in \Sec{Matrix_elements}. In Fig.~\ref{fig:Ph}, we show the Feynman diagrams for the single graviton production process $\phi\phi \to\phi \phi h$ at order $\Ott$.

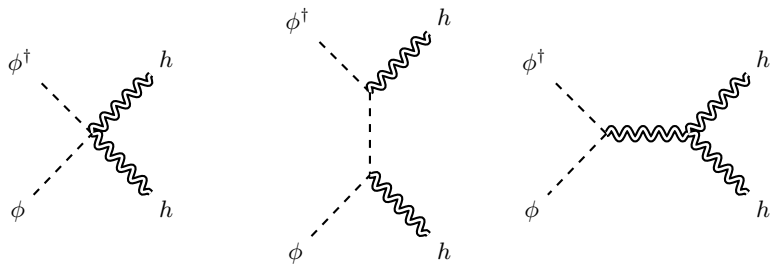
\begin{figure}[t]
	\centering
	\begin{subfigure}[c]{.2\textwidth}
	\begin{tikzpicture}
	\begin{feynman}
	\vertex (a) {\(\phi^\dagger \)};
	\vertex [below right=1.35cm of a] (b) ;
	\vertex [above right=1.1cm of b] (c){\(h\)};
	\vertex [below right=1.1cm of b] (d){\(h\)};
	\vertex [below left=1.1cm of b] (e){\(\phi\)} ;	
	\diagram* {
		(b) -- [scalar,thick] (a), (b) -- [thick, double distance=1pt, decoration={snake, amplitude=2pt, segment length=6pt}, decorate] (c),
		(b) -- [thick, double distance=1pt, decoration={snake, amplitude=2pt, segment length=6pt}, decorate] (d), (e) -- [scalar,thick] (b)
	};
	\end{feynman}
	\end{tikzpicture}
	\end{subfigure}
	\begin{subfigure}[c]{.2\textwidth}
	\begin{tikzpicture}
	\begin{feynman}
	\vertex (a) {\(\phi^\dagger \)};
	\vertex [below right=1.35cm of a] (b) ;
	\vertex [above right=1.1cm of b] (c){\(h\)};
	\vertex [below=1.1cm of b] (d);
	\vertex [below right=1.1cm of d] (e) {\(h\)};	
	\vertex [below left=1.1cm of d] (f){\(\phi\)} ;
	\diagram* {
		(b) -- [scalar,thick] (a), (b) -- [thick, double distance=1pt, decoration={snake, amplitude=2pt, segment length=6pt}, decorate] (c), 
		(b) -- [scalar,thick] (d), (d) -- [thick, double distance=1pt, decoration={snake, amplitude=2pt, segment length=6pt}, decorate] (e), (f) -- [scalar,thick] (d)
	};
	\end{feynman}
	\end{tikzpicture}
	\end{subfigure}
	\begin{subfigure}[c]{.2\textwidth}
	\begin{tikzpicture}
	\begin{feynman}
	\vertex (a) {\(\phi^\dagger \)};
	\vertex [below right=1.35cm of a] (b) ;
	\vertex [right=1.1cm of b] (c);
	\vertex [above right=1.1cm of c] (d){\(h\)};
	\vertex [below right=1.1cm of c] (e){\(h\)} ;	
	\vertex [below left=1.1cm of b] (f){\(\phi\)} ;	
	\diagram* {
		(b) -- [scalar,thick] (a), (b) -- [thick, double distance=1pt, decoration={snake, amplitude=2pt, segment length=6pt}, decorate] (c),
		(c) -- [thick, double distance=1pt, decoration={snake, amplitude=2pt, segment length=6pt}, decorate] (d), (c) -- [thick, double distance=1pt, decoration={snake, amplitude=2pt, segment length=6pt}, decorate] (e),
		(b) -- [scalar,thick] (f)
	};
	\end{feynman}
	\end{tikzpicture}
	\end{subfigure}
	\caption{Lowest order Feynman diagrams for the process $\phi^\dagger\phi\to h h$. The matrix element squared is of the order $\mathcal{O}\left(\kappa^4\right)$. There is one additional diagram that is not shown since it can be obtained by crossing the final state gravitons in the second diagram.}
	\label{fig:Ph_phi_phi_h_h}
\end{figure}

The contributions to $\dfh$ at order $\Ozf$  are sourced by processes that have two gravitons in the final state.
These processes can be relevant in high-temperature early universe scenarios since the dimensionless expansion parameter is $T/\Mp$ which can be relatively large if the temperature is close to the Planck mass. The explicit form of $\dfh^{(0,4)}$ is:
\begin{eqnarray}
	\dfh^{(0,4)}=\int_{\phi^\dagger\phi\to h h}^{(0,4)} \nb \nb.
	\label{eq:dfh40}
\end{eqnarray}
Processes with a graviton in the initial state do not contribute, since the initial state graviton always comes with a factor $\fh$ which makes the whole term of higher order. The same holds for  final 
state $1+\fh$ amplification factors.
The corresponding Feynman diagrams that contribute to two graviton production are shown in \Fig{Ph_phi_phi_h_h}. In \Sec{Matrix_elements} we calculate  these diagrams.

In the following we discuss how one could extend our calculation of $\fh$ to higher orders. In particular we point out the limitations and challenges that one would face.
The evolution \Eq{evolution_eq_fphi} for the graviton distribution function can definitively be used without problems at lowest order in perturbation theory, i.e., in our case these are the $2\to 2$ and $2\leftrightarrow 3$ processes that are of the order $\Ozf$ and $\Ott$ respectively. These lowest order terms have \emph{real corrections}, and \emph{virtual corrections}. Virtual corrections are loop correction,
while real corrections come from tree-level processes with extra initial-
or final-state particles.
If they are finite, the latter are easily incorporated into the Boltzmann equation formalism, i.e., \ Eq.~\eqref{eq:Gh}. 
Incorporating the former, on the other hand, is not straightforward as the matrix elements squared contain not
only standard vacuum fluctuations but also statistical fluctuations which, in turn,
depend on the distribution functions themselves~\cite{Laine:2016hma}. 
Furthermore, while real
and renormalized virtual corrections might separately be finite in a standalone scalar
theory, in more complex systems such as the scalar theory coupled to 
gravity or gauge theories they are in general not finite, with infrared (IR) divergences
canceling between the two, as in the case of the 
Kinoshita--Lee--Nauenberg theorem~\cite{Kinoshita:1962ur,Lee:1964is}. 
In conclusion it is a challenging task to incorporate higher order effects with the Boltzmann-like approach since it is only possible to incorporate the finite higher order effects.

\begin{figure}[t]
	\centering
	\begin{tikzpicture}
	\begin{feynman}
	\vertex (a) {\(\phi^\dagger \)};
	\vertex [below right=1.1cm of a] (b);
	\vertex [below right=0.9cm of b] (c);
	\vertex [above right=0.9cm of c] (d);
	\vertex [above right=0.9cm of d] (e) {\(\phi^\dagger \)};
	\vertex [below right=0.9cm of c] (f);
	\vertex [right=0.9cm of f] (g) {\(h \)};
	\vertex [below right=0.9cm of f] (h) {\(\phi \)};
	\vertex [below left=0.9cm of c] (i);
	\vertex [below left=0.9cm of i] (j) {\(\phi \)};
	\diagram* {
		(a) -- [scalar,thick] (b), (b) -- [scalar,thick] (c),
		(c) -- [scalar,thick] (d), (d) -- [scalar,thick] (e),
		(c) -- [scalar,thick] (f), (f) -- [thick, double distance=1pt, decoration={snake, amplitude=2pt, segment length=6pt}, decorate] (g),
		(f) -- [scalar,thick]  (h),
		(c) -- [scalar,thick] (i), (i) -- [scalar,thick] (j),
	};
	\end{feynman}
	\end{tikzpicture}
    ~~~
    \begin{tikzpicture}
	\begin{feynman}
	\vertex (a) {\(\phi^\dagger \)};
	\vertex [below right=1.1cm of a] (b);
	\vertex [below right=0.9cm of b] (c);
	\vertex [above right=0.9cm of c] (d);
	\vertex [above right=0.9cm of d] (e) {\(\phi^\dagger \)};
	\vertex [below right=0.9cm of c] (f);
	\vertex [right=0.9cm of f] (g) {\(h \)};
	\vertex [below right=0.9cm of f] (h) {\(\phi \)};
	\vertex [below left=0.9cm of c] (i);
	\vertex [below left=0.9cm of i] (j) {\(\phi \)};
	\diagram* {
		(a) -- [scalar,thick] (b), (b) -- [scalar,thick] (c),
		(b) -- [thick, double distance=1pt, decoration={snake, amplitude=2pt, segment length=6pt}, decorate] (d),
		(c) -- [scalar,thick] (d), (d) -- [scalar,thick] (e),
		(c) -- [scalar,thick] (f), (f) -- [thick, double distance=1pt, decoration={snake, amplitude=2pt, segment length=6pt}, decorate] (g),
		(f) -- [scalar,thick]  (h),
		(c) -- [scalar,thick] (i), (i) -- [scalar,thick] (j),
	};
	\end{feynman}
	\end{tikzpicture}
	~~~
		\begin{tikzpicture}
	\begin{feynman}
	\vertex (a) {\(\phi^\dagger \)};
	\vertex [below right=1.1cm of a] (b);
	\vertex [below right=0.9cm of b] (c);
	\vertex [above right=0.9cm of c] (d);
	\vertex [above right=0.9cm of d] (e) {\(\phi^\dagger \)};
	\vertex [right=0.9cm of d] (k) {\(h \)};
	\vertex [below right=0.9cm of c] (f);
	\vertex [right=0.9cm of f] (g) {\(h \)};
	\vertex [below right=0.9cm of f] (h) {\(\phi \)};
	\vertex [below left=0.9cm of c] (i);
	\vertex [below left=0.9cm of i] (j) {\(\phi \)};
	\diagram* {
		(a) -- [scalar,thick] (b), (b) -- [scalar,thick] (c),
		(c) -- [scalar,thick] (d), (d) -- [scalar,thick] (e),
		(d) -- [thick, double distance=1pt, decoration={snake, amplitude=2pt, segment length=6pt}, decorate] (k),
		(c) -- [scalar,thick] (f), (f) -- [thick, double distance=1pt, decoration={snake, amplitude=2pt, segment length=6pt}, decorate] (g),
		(f) -- [scalar,thick]  (h),
		(c) -- [scalar,thick] (i), (i) -- [scalar,thick] (j),
	};
	\end{feynman}
	\end{tikzpicture}
	\caption{The interference of the left and center diagrams yields a term $\Otf$ in the  matrix element squared. The right diagram denotes a real correction that is on the same order when it's squared.}
	\label{fig:Ph_qg_effects}
\end{figure}
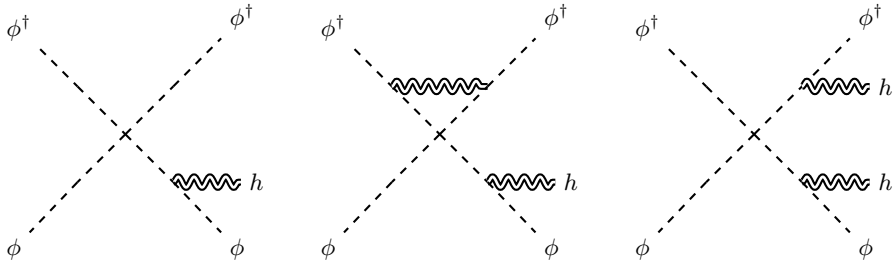

Virtual gravitons arise already at order $\Ozf$ in the graviton distribution function, cf. \Fig{Ph_phi_phi_h_h}.
Quantum gravity effects start to play a role at order $\Otf$, since at this order diagrams with graviton loops exist. The three diagrams that we show in  \Fig{Ph_qg_effects} are of order $\Ooo$, $\Ooth$ and 
$\mathcal{O}\left(\lambda^1 \kappa^2\right)$ respectively. The interference term of the first two diagrams is of order $\Otf$ and is the first virtual correction
involving loops of gravitons. Conversely, the square of the third diagram is
$\mathcal{O}\left(\lambda^2 \kappa^4\right)$ and is part of the real corrections
at that order. This further exemplifies the challenge in going beyond 
leading order: the real corrections can be dealt with in a Boltzmann 
form in a rather straightforward way, while the virtual corrections
cannot, as their matrix element squared 
will depend in non-trivial ways on the statistical factors --- see Ref.~\cite{Laine:2022ner} for a recent work on this problem in a non-gravitational
setting.

In an alternative way, one could systematically study quantum effects in kinetic theory from first principles using the Wigner-function formalism. By performing an expansion in $\hbar$ of the Wigner function, one can in principle derive quantum corrections to the classical Boltzmann equation, see, e.g., Refs~\cite{Weickgenannt:2019dks,Weickgenannt:2020aaf,Yang:2020hri,Weickgenannt:2021cuo,Sheng:2021kfc}. The development of such quantum kinetic theory is left for future studies.

Along the lines that we discussed before, we want to mention that it is possible to calculate $\dfh^{(\text{all},2)}$ in a full quantum picture. 
As shown in Ref.~\cite{Ghiglieri:2020mhm} for the specific case 
of GWs and more generally in Ref.~\cite{Bodeker:2015exa} for
any state that is feebly coupled to a thermal bath, 
the Boltzmann-equation based approach that we use here agrees, at $\Ott$,
with 
the thermal-field-theoretical approach of
\emph{production and equilibration rates}.
Namely, for GWs one has  $\dfh^{(\text{all},2)}(k)=\Gamma(k)\big[\nb(k)-\fh^{(\text{all},2)}(k)\big]$,~\cite{Ghiglieri:2015nfa} where the production/equilibration rate $\Gamma(k)$
is proportional to $\kappa^2$ times the imaginary part of the
retarded two-point function of the $T_{12}$ component 
of the energy-momentum tensor 
of the equilibrium particles, i.e., in our case, the scalars.
This formalism defines the single-graviton
production rate to all orders in $\lambda$.
Within this formalism, higher orders in $\lambda$
naturally incorporate both real and virtual corrections, 
without the issues that would plague direct attempts 
in the Boltzmann-like approach.
However note that, in principle, we do not want to go to higher orders in $\lambda$ but to higher orders in $\kappa$ to identify quantum gravity effects. The discussed thermal-field-theoretical approach is not suited for this and new strategies have to be developed for a full quantum treatment of the graviton production rate.

Finally let us discuss \emph{back-reaction} effects, which can be incorporated into the Boltzmann-like formalism.
If we stay at order $\Ozf$ and go to non-zero order in $\lambda$ we can identify back-reaction effects.
These appear at lowest order at $\mathcal{O}(\kappa^4\lambda^4)$. 
The graviton production rate at this order contains the following back-reaction terms:
\begin{eqnarray}
	\dfh^{(4,4)}(k)= \int_{\phi\phi\to\phi\phi h}^{(2,2)} \fp^{(2,2)} \nb (1+\nb)(1+\nb)+\int_{\phi\phi^\dagger\to h h}^{(0,4)}\fp^{(4,0)}\nb+\cdots,
	\label{eq:ExampleBackReaction}
\end{eqnarray}
where the dots above stand for other terms that we have omitted here.
We call the terms in \Eq{ExampleBackReaction} back-reaction terms since the small corrections on top of the Bose Einstein distribution, $\fp^{(2,2)}$ and $\fp^{(0,4)}$, appear in the phase space integral. Back-reaction effects also appear in the $\dfp$ rate. 

We further note that the RHS in \Eqs{dfh22}{dfh40} are time independent. Therefore $\fh^{(2,2)}$ and $\fh^{(0,4)}$, are linear in time and the back-reaction rates, i.e. $\dfh^{(4,4)}$, are linear in time. From this follows that $\fh^{(4,4)}$ has a quadratic time dependence.

\section{Matrix elements and Phase space integrals}\label{sec:Matrix_elements}
In this section we calculate the matrix elements squared for graviton production at order $\Ott$ and $\Ozf$. 
Let us start with the $\Ott$ component. As argued 
previously, it arises from  $2\to 3$
 and  $3\to 2$ processes. 
The corresponding expressions for the distribution functions are:
\begin{align}
  \label{kin_thy23}
  \dfh^{2\to 3}(k)=&\frac{1}{16k}
  \int \! {\rm d}\Omega^{ }_{2\to3} \sum_{ abcd}
  \Bigl\vert\mathcal{M}^{ ab}_{ cdh}
  (\vp_1',\vp_2';\vp_1,\vp_2,\vk)\Bigr\vert^2 
  \nB^{ }(p'_1)\,\nB^{ }(p'_2)\,[1+ \nB^{ }(p_1)]\,
  [1 + \nB^{ }(p_2)]\,,\\
   \dfh^{3\to 2}(k)=&\frac{1}{24k}
    \int \! {\rm d}\Omega^{ }_{3\to2} \sum_{abcd}
    \Bigl\vert\mathcal{M}^{abc}_{dh}
    (\vp_1',\vp_2',\vp_3';\vp_1,\vk)\Bigr\vert^2 
    \nB^{ }(p_1')\,\nB^{ }(p_2')\,\nB^{ }(p_3')\,[1+ \nB^{ }(p_1)]
    \;,
    \label{kin_thy32}\\
    \dfh^{(2,2)}(k)=&\dfh^{2\to 3}(k)+\dfh^{3\to 2}(k),
    \label{kin_thylambdasq}
\end{align}
where at order $\Ott$ there is no $\fh$ on the right-hand side.
The sums run over all $abcd$ 
scalar and antiscalar degrees of freedom  
and thus over all ${ab}\to {cd h}$ and  ${abc}\to {d h}$ processes,
with  $\h$ denoting the graviton. The quantities 
$\vert\mathcal{M}^{ab}_{ cdh}(\vp_1',\vp_2';\vp_1,\vp_2,\vk)
\vert^2$  and
$\vert\mathcal{M}^{abc}_{ dh}(\vp_1',\vp_2',\vp_1';\vp_1,\vk)
\vert^2$ 
are the corresponding matrix elements squared
summed over the graviton polarizations.
For $k\sim  T$  the contribution of the  thermal mass $m_{\phi\,T}=\sqrt{\lambda/12} T$ is suppressed, 
so the external states can be considered massless.
The prefactor $1/(16k)$ is a combination of $1/(2k)$ from 
the phase space measure, 
$1/2$ for the graviton polarization degeneracy, and $1/(2!)^2$ for 
the symmetry factors for identical initial and final state particles.
In the cases where $ a\ne  b$ or $ c\ne  d$
the sum over $abcd$ counts the process two times and compensates for this factor.
Similarly, $1/(24k)$ is a combination of $1/(2k)$ from 
the phase space measure, 
$1/2$ for the graviton polarization degeneracy, and $1/(3!)$ for 
the symmetry factors for identical initial state particles. 

The phase spaces can be read off from Eq.~\eqref{eq:phasespace}.
For the matrix element squared we used the automated pipeline introduced
in Ref.~\cite{Ghiglieri:2020mhm}.
We first used \textsc{FeynRules}~\cite{Alloul:2013bka} to derive
Feynman rules for the  Lagrangian in Eq.~\eqref{eq:Lagrangian_linearized_gravits}.
Using the appropriate interface~\cite{Christensen:2009jx},
\textsc{FeynRules} generates a \textit{model file}
for \textsc{Feyn\-Arts}~\cite{hep-ph/0012260}.
This package and its companion \textsc{FormCalc}~\cite{1604.04611} 
were then used to generate, evaluate
and square all amplitudes. 
Tensor boson polarization sums had
to be implemented following the method discussed in Ref.~\cite{Ghiglieri:2020mhm}.
For $\phi\phi^\dagger\to\phi\phi^\dagger \h$, the results is
\begin{align}
    \left\vert\mathcal{M}^{\phi\phi^\dagger}_{\phi\phi^\dagger \h}(\vp_1',\vp_2';\vp_1,\vp_2,\vk)
\right\vert^2
=&
\frac{\kappa^2\lambda^2}{2}\bigg[
\frac{(P_1'\cdot P_2')^2}{P_1'\cdot K\,P_2'\cdot K }  +
\frac{(P_1\cdot P_2)^2}{P_1\cdot K\,P_2\cdot K }-
\frac{(P_1'\cdot P_2)^2}{P_1'\cdot K\,P_2\cdot K }-
\frac{(P_1\cdot P_2')^2}{P_1\cdot K\,P_2'\cdot K }\nonumber \\
&-
\frac{(P_1'\cdot P_1)^2}{P_1'\cdot K\,P_1\cdot K }-
\frac{(P_2'\cdot P_2)^2}{P_2'\cdot K\,P_2\cdot K }-2
\bigg].\label{23mat}
\end{align}
Equation~\eqref{23mat} arises from  diagrams such as the ones in Fig.~\ref{fig:Ph}.
The four structures in the
denominator, e.g. $P_1'\cdot K$, correspond to the
propagator of the virtual, intermediate scalar connecting
the $\phi\phi \h$ vertex with the $\phi^4$ one. 
It is reassuring to see that this matrix element squared is finite even when one of the scalar products in the denominators vanishes. 
For example, the term $P_1'\cdot K$ can vanish either for $p_1'\to 0$ or when $\vp_1'$ is parallel to $\vk$.
In the former case the powers of $p_1'$ at the numerator immediately remove
the divergence, and similarly the phase space is free of endpoint divergences
for $p_1'\to 0$, even in the presence of Bose enhancement 
($n_\mathrm{B}(p_1'\to 0)\approx T/p_1'$). 
In the collinear case, $\vp_1'\parallel \vk$, one can show that 
the sum of the divergent terms is finite.

Let us now discuss the terms that are obtained by crossing. By crossing 
an initial state $\phi$ ($ \phi^\dagger$) to the 
final state one obtains a  $\phi^\dagger$ ($ \phi$). 
Hence one finds
\begin{equation}
    \left\vert\mathcal{M}^{\phi\phi^\dagger}_{\phi\phi^\dagger \h}(\vp_1',\vp_2';\vp_1,\vp_2,\vk)
    \right\vert^2=
    \left\vert\mathcal{M}^{\phi\phi}_{\phi\phi \h}(\vp_1',\vp_2';\vp_1,\vp_2,\vk)
    \right\vert^2
    =    \left\vert\mathcal{M}^{\phi^\dagger\phi^\dagger}_{\phi^\dagger\phi^\dagger \h}(\vp_1',\vp_2';\vp_1,\vp_2,\vk)
    \right\vert^2.
\end{equation}
Furthermore, $\left\vert\mathcal{M}^{\phi\phi^\dagger}_{\phi\phi^\dagger \h}\right\vert^2$
is symmetric under permutations within the initial and final states of the $\phi$ 
and $\phi^\dagger$.
Hence, in the sum over $abcd$ of Eq.~\eqref{kin_thy23}, it is counted
four times, whereas the $\phi$-only or $\phi^\dagger$-only processes are
counted once.

The $3\to 2$ matrix elements squared can be obtained by crossing 
Eq.~\eqref{23mat}, too. For instance, if we cross the final-state
 $\phi^\dagger$ with momentum $P_2$ into an initial-state $\phi$ 
 with momentum $P_3'$ we have
\begin{align}
    \left\vert\mathcal{M}^{\phi\phi^\dagger\phi}_{\phi \h}
    (\vp_1',\vp_2',\vp_3';\vp_1,\vk)
\right\vert^2
=&
\frac{\kappa^2\lambda^2}{2}\bigg[
\frac{(P_1'\cdot P_2')^2}{P_1'\cdot K\,P_2'\cdot K }  +
\frac{(P_1'\cdot P_3')^2}{P_1'\cdot K\,P_3'\cdot K }
+
\frac{(P_2'\cdot P_3')^2}{P_2'\cdot K\,P_3'\cdot K }\nonumber \\
&-
\frac{(P_1\cdot P_1')^2}{P_1\cdot K\,P_1'\cdot K }-
\frac{(P_1\cdot P_2')^2}{P_1\cdot K\,P_2'\cdot K }-
\frac{(P_1\cdot P_3')^2}{P_1\cdot K\,P_3'\cdot K }-2
\bigg].\label{32mat}
\end{align}
By further crossing one finds
\begin{equation}
    \left\vert\mathcal{M}^{\phi^\dagger\phi^\dagger\phi}_{\phi^\dagger \h}
    (\vp_1',\vp_2',\vp_3';\vp_1,\vk)
\right\vert^2=     \left\vert\mathcal{M}^{\phi\phi^\dagger\phi}_{\phi \h}
(\vp_1',\vp_2',\vp_3';\vp_1,\vk)
\right\vert^2.
\end{equation}
The processes with two $\phi$ and with two $\phi^\dagger$ in the 
initial state are counted three times each in the sum over $abcd$.
Putting everything together we then find
\begin{align}
\label{kin_thy23_expl}
       \dfh^{2\to 3}(k)=&\frac{3}{8k}
  \int \! {\rm d}\Omega^{ }_{2\to3} 
  \Bigl\vert\mathcal{M}^{\phi\phi}_{\phi\phi \h}
  (\vp_1',\vp_2';\vp_1,\vp_2,\vk)\Bigr\vert^2 
  \nB^{ }(p_1')\,\nB^{ }(p_2')\,[1+ \nB^{ }(p_1)]\,
  [1 + \nB^{ }(p_2)]\,,\\
   \dfh^{3\to 2}(k)= &\frac{1}{4k}
    \int \! {\rm d}\Omega^{ }_{3\to2} 
    \Bigl\vert\mathcal{M}^{\phi\phi^\dagger\phi}_{\phi \h}
    (\vp_1',\vp_2',\vp_3';\vp_1,\vk)\Bigr\vert^2 
    \nB^{ }(p_1')\,\nB^{ }(p_2')\,\nB^{ }(p_3')\,[1+ \nB^{ }(p_1)]
    \;.\label{kin_thy32expl}
\end{align}
Details on the seven-dimensional numerical integration of the phase space can be found in \App{integrate}.

The $\Ozf$ component is sourced by $2\to 2$ processes:
\begin{equation}
  \label{eq:kin_thy22}
\dfh^{(0,4)}(k)=\frac{1}{8k}
  \int \! {\rm d}\Omega^{ }_{2\to2} \sum_{\rm ab}
  \Bigl\vert\mathcal{M}^{ab}_{\h\h}
  (\vp_1',\vp_2';\vk,\vk_2)\Bigr\vert^2 
  \nB^{ }(p_1')\,\nB^{ }(p_2')
  \;,
\end{equation}
where $1/(8k)$ is the product of the $1/(2k)$ from the Lorentz phase space measure,
a factor of $1/2$ for the two polarizations of the graviton and
a factor of $1/(2!)$ for possible identical initial state particles.

The matrix element squared arises from four diagrams.
Three of them are shown in Fig.~\ref{fig:Ph_phi_phi_h_h}, and the fourth
comes from the $u$-channel analog of the  second diagram. We have computed the matrix element squared by using the 
previously described \textsc{FeynRules}, \textsc{FeynArts} and 
\textsc{FormCalc} machinery:
\begin{equation}
    \Bigl\vert\mathcal{M}^{\phi\phi^\dagger}_{\h\h}
  (\vp_1',\vp_2';\vk,\vk_2)\Bigr\vert^2 
  =
  \frac{\kappa^4}{8} \frac{t^2u^2}{s^2},
\end{equation}
where $s$, $t$ and $u$ are the standard Mandelstam invariants.
Note that the matrix element squared can also be
extracted from Refs.~\cite{Holstein:2006bh,Bjerrum-Bohr:2014lea}\footnote{
These papers show that the amplitudes for
graviton production factorize into simple products of photon amplitudes 
times kinematic factors. The massless limit of Eq.~(21) of Ref.~\cite{Holstein:2006bh}
gives the $\phi\gamma\to\phi\gamma$ scalar Compton amplitudes in 
the helicity basis. Eq.~(62) of
the same paper then gives the matrix elements squared as 
the fourth power of these amplitudes, multiplied by 
the second power of the kinematic factor $F$, given in Eq.~(61). The matrix elements squared for the 
two polarizations are identical and thus trivially summed.
Finally, crossing symmetry relates the $\phi \h \to \phi \h$
matrix element squared to the $\phi^\dagger\phi\to h h$ matrix element squared} and we have checked that our results agree.
Accounting for a factor of $2$ from the sum in ~\Eq{kin_thy22} yields:
\begin{equation}
  \label{eq:kin_thy22_expl}
\dfh^{(0,4)}(k)=\frac{1}{4k}
  \int \! {\rm d}\Omega^{ }_{2\to2} \Bigl\vert\mathcal{M}^{\phi\phi^\dagger}_{\h\h}
  (\vp_1',\vp_2';\vk,\vk_2)\Bigr\vert^2 
  \nB^{ }(p_1')\,\nB^{ }(p_2')
  \;.
\end{equation}
We refer again to \App{integrate} for details on the 
reduction of the phase space integration to a two-dimensional integral that we evaluate
numerically.

\begin{figure}
	\centering
	\includegraphics[width=0.5\textwidth]{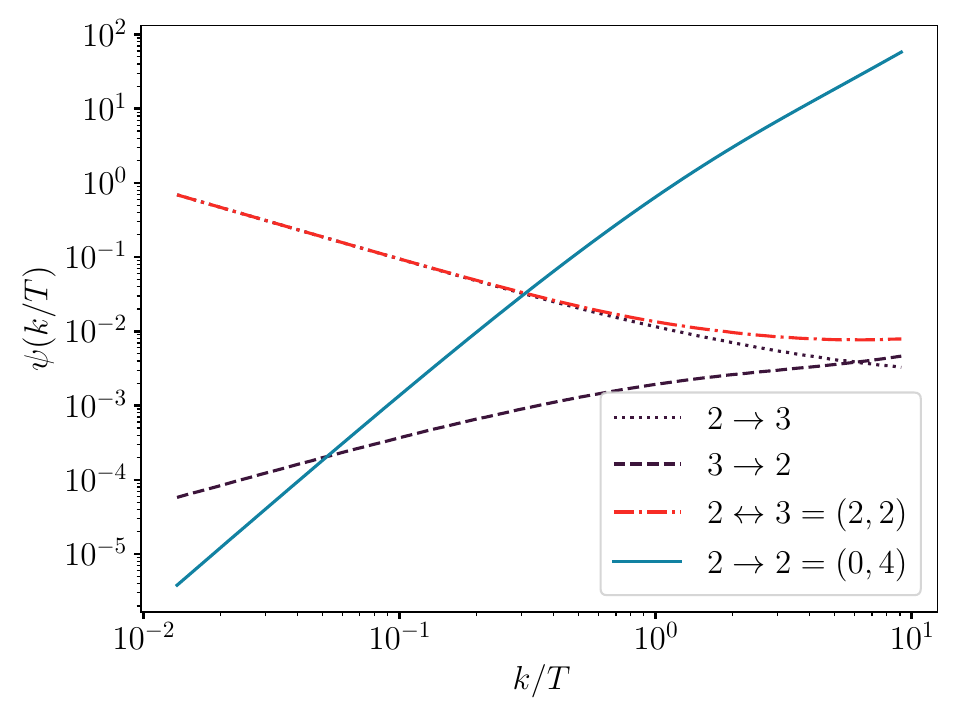}
	\caption{We plot the $\psi$-functions that have been obtained by numerically integrating the phase space integrals. The results are shown for the $2\to 3$, $3\to 2$ and $2\to 2$ processes respectively. The sum of the $2\to 3$ and $3\to 2$ contributions is denoted as $2\leftrightarrow 3$. $2\leftrightarrow 3$ processes come from single graviton production processes that have a lowest order matrix element squared of the order $\Ott=(2,2)$. $2\to 2$ processes arise in graviton pair production processes with a lowest order matrix element squared of the order $\Ozf$.}
	\label{fig:fig_phi_functions}
\end{figure}

The production rates for $(i,j)=(2,2)$ and $(i,j)=(0,4)$ can be written in a compact form as:
\begin{eqnarray}
	\dfh^{(i,j)}(k)=\frac{1}{2k}\,T^2\,\left(\frac{T}{\Mp}\right)^j\, \lambda^i\, \nb\left(\frac{k}{T}\right)\, \psi^{(i,j)}\left(\frac{k}{T}\right),
	\label{eq:fh_form}
\end{eqnarray}
where we have defined dimensionless $\psi$-functions and used the convention that $\nb(x)=1/(e^{x}-1)$ if the argument of $\nb$ is dimensionless. The $\psi$-functions are shown in \Fig{fig_phi_functions} for the $2\to 3$ (dotted black) and $3\to 2$ (dashed black) processes. We also show the sum of the $2\to 3$ and $3\to 2$ processes which is labeled as $2\leftrightarrow 3$ and shown as a red dot-dashed line. The $2\to 2$ processes are shown as a solid blue line. Note that while the $\psi$-function for the $2\leftrightarrow 3$ processes has only a relatively mild $k/T$ dependence around $k/T\simeq 1$ this is not the case for the $2\to 2$ processes. 
In \App{integrate} we have derived an asymptotic form for $\dfh^{(0,4)}$, i.e., $\psi^{(0,4)}$ in the limit $k>T$. We find $\psi^{(0,4)}=32/(15 \pi) (k/T)^2$ for $k>T$ what confirms the results in \Fig{fig_phi_functions}.
$\psi^{(2,2)}$ and $\psi^{(0,4)}$ differ in their functional form due to the fact that the underlying phase space structure and  matrix elements are different for single and graviton pair production. The single gravitons are produced in $2\leftrightarrow 3$ processes, while graviton pair production is a $2\to 2$ process. 
In addition, the non-renormalizable nature of gravity leads to a matrix element squared for single graviton production that is proportional to $\kappa^2\sim \frac{1}{\Mp^2}$
times an expression with dimensions $({\rm energy})^2$. Similarly the matrix element squared for graviton pair production 
is proportional to $\kappa^4\sim \frac{1}{\Mp^4}$ times an expression with dimensions $({\rm energy})^4$.
In the case of graviton pair production
the $2\to 2$ phase space integration for large $k$ translates this into
a quadratic dependence on momentum, which explains the form of $\psi^{(0,4)}$ for large $k$.

As \Fig{fig_phi_functions} shows, 
the $2\to 3$ contribution to $\dfh^{(2,2)}$
leads to 
$\dfh^{(2,2)}\propto k^{-3}$ for $k\ll T$, which implies a naively IR-divergent contribution to 
the number density of gravitons ($n_{\rm gw}\propto \int dk k^2 \fh$) 
and a finite but enhanced contribution
to the energy density $\rho_{\rm gw}\propto \int dk k^3 \fh$ from the IR domain $k\ll T$. 
This  IR contribution
is an artifact of treating the external and intermediate scalar states as massless.
If we would include their thermal mass $m_{\phi\,T}$,
then such a behavior would go away, as the matrix element would no longer
behave like $1/k^2$ at small $k$. Scalars have 
the nice property that their thermal mass behaves like an ordinary local
 mass term in the Lagrangian, unlike gauge fields and fermions. 
This in turn would make thermal mass
 resummation relatively straightforward. For this paper
we limit ourselves to unresummed (massless) results, and consider 
the result for $\dfh$ 
to be a proper leading-order determination in the regime $k\gtrsim m_{\phi\,T}= \sqrt{\lambda/12}\, T$.
Furthermore, for smaller $k$, $k\ll \lambda^2 T$,
the quasi-particle description breaks 
down completely and gravitational waves are sourced from hydrodynamic fluctuations~\cite{Ghiglieri:2015nfa}.

\section{Gravitational wave spectrum}\label{sec:embeeding_in_cosmology}

In this section we embed the graviton production rate into cosmological evolution. Our main assumption is that, after the hot Big Bang, a thermal plasma of $\phi$-particles with temperature $\Tmax$ is present.  Throughout the expansion of the universe the thermal plasma produces GWs.
From the definition of \Eq{f_h_definition} it follows that the GW differential energy density is $d\rhoGW(t,k)=2k \fh(t,k)\frac{d^3k}{(2\pi)^3}$, where a flat space-time metric has been assumed, and the factor of $2$ takes the two polarization states into account which contribute to the energy density.
We can rewrite the equation for the energy density as
$\frac{d\rhoGW}{dt\, d{\rm ln} k}=\frac{k^4 }{\pi^2}\,\dfh$\,.
Generalizing to an expanding universe, the GW energy density evolves as~\cite{Ghiglieri:2015nfa}
\begin{eqnarray}
(\partial_t+4H)\rhoGW(t)=\int \frac{d^3k}{(2\pi)^3} R(t,k),
\label{eq:energy_density_evolution}
\end{eqnarray}
where $H$ is the Hubble parameter and we have defined $R(t,k)\equiv\,2 \,k\,\dfh(t,k)$. Note that for the $\Ott$ and $\Ozf$ contributions, which we shall discuss here, $\dfh$ has no explicit time dependence. We will therefore treat $\dfh$ without explicit time dependence in the following derivation. 
Now that we consider a thermal plasma in an expanding universe, the temperature decreases over time. Therefore we have an implicit time/temperature dependence.

We further note that, in a radiation-dominated universe,
\begin{equation}
    \label{hubrad}
    H=\sqrt{\frac{8\pi \rho(T)}{3}}\frac{1}{\Mp},
\end{equation}
with $\rho(T)=\gsr(T)\pi^2T^4/30$ and $\gsr(T)=2$ is the effective number of energy density degrees of freedom. 
The scalars remain in thermal equilibrium\footnote{
By this we mean that the zeroth order term of $\fp$ is a massless Bose--Einstein distribution with the current temperature.
} as long as their 
interaction rate, which is on the order $\lambda^2 T$, 
is at least as fast as the Hubble rate $H\sim T^2/\Mp$.
Therefore we obtain the equilibrium condition, $\lambda^2 \gtrsim  T/\Mp$, which has to be understood as an order of magnitude estimate.

We can integrate \Eq{energy_density_evolution}:
\begin{eqnarray}
\frac{\rho_{\rm gw}(t_1)}{s^{4/3}(t_1)}-\frac{\rho_{\rm gw}(t_0)}{s^{4/3}(t_0)}=\int_{t_0}^{t_1}dt\,\frac{1}{s^{4/3}(t)}\int \frac{d^3k}{(2\pi)^3}\, R(T(t),k),
\label{eq:energy_density_integral_form}
\end{eqnarray}
where we have used that the entropy density $s$ fulfills $\dot{s}+3Hs=0$ and we have made the temperature/time dependence explicit.
We assume that at the beginning of the GW production no GWs are present, i.e., $\rho_{\rm gw}(t_0)=0$. At $t_0$ the thermal plasma was first in thermal equilibrium and had its maximum temperature $\Tmax$. The time $t_1$ is the time when the mass of the scalar field cannot be neglected anymore. The temperature corresponding to the time $t_1$ is referred to as $\Tphi$ in the following. 
In \Eq{energy_density_integral_form} we can only integrate to $t_1$
since the production rates that we have calculated are only valid for temperatures above $\Tphi$. 
The time integral in \Eq{energy_density_integral_form} can be transformed into an integral over the temperature by using the relation~\cite{Laine:2015kra}:
\begin{eqnarray}
\frac{dT}{dt}=-\sqrt{\frac{4\pi^3}{45}}\gsr(T)^{\frac{1}{2}}\frac{g_{*s}(T)}{\gsc(T)}\frac{T^3}{\Mp},
\end{eqnarray}
where $\gss$ and $\gsc$ are the effective degrees of freedom for the entropy density and heat capacity, which are defined as $s(T)\equiv\gss (T)\frac{2\pi^2}{45}T^3$ and $c(T)\equiv\gsc (T) \frac{2\pi^2}{15}T^3$.
In the following we use the assumption of isotropy under which we can simplify the $d^3k$ integral: $\int d^3k\, R(T,k)=4\pi \int dk\, k^2 R(T,k)=4\pi \int d\text{ln}(k) k^3 R(T,k)$.
From \Eq{energy_density_integral_form} we can then read off the GW energy density per logarithmic momentum interval at $\Tphi$, normalized to the total energy density: $\OmGW=\frac{1}{\rho}\frac{d\rhoGW}{d{\rm ln}k}$. Redshifting all corresponding quantities to today~\cite{Ringwald:2020ist} yields:
\begin{eqnarray}
h^2_0\OmGW(\fg)&=&\frac{15\sqrt{45}}{4\pi^{11/2}}\Mp\, \gss(T_{\rm today})^{1/3}\,h^2_0\Omega_{\gamma}\left(\frac{2\pi\fg}{T_{\rm today}}\right)^3\,\times \nonumber\\
&~&\times\int_{\Tphi}^{\Tmax} dT \frac{1}{T^4} \frac{\gsc(T)}{\gsr(T)^{1/2}\gss(T)^{4/3}}R\left(T,T\frac{2\pi\fg}{T_{\rm today}}\left(\frac{\gss(T)}{\gss(T_{\rm today})}\right)^{1/3}\right),
\label{eq:OmegaGWToday}
\end{eqnarray}
where $\fg$ is the current day GW frequency, $h^2_0\Omega_\gamma=2.473\times 10^{-5}$ is the present fractional photon energy density, $h^2_0$ a factor that eliminates the experimental uncertainty that is coming from measurements of the Hubble constant, $T_{\rm today}=2.7254$ K the current day temperature~\cite{ParticleDataGroup:2022pth} and $\gss(T_{\rm today})=3.931$ are the effective entropy degrees of freedom today~\cite{Saikawa:2018rcs}.

With the parametric form of $\dfh$ from \Eq{fh_form} we can write $R$ as:
\begin{eqnarray}
R(T,k)=2k\dfh(T,k)= T^2 \nb\left(\frac{k}{T}\right) \frac{T^2}{\Mp^2}\left(\lambda^2\psi^{(2,2)}\left(\frac{k}{T}\right)+\frac{T^2}{\Mp^2}\psi^{(0,4)}\left(\frac{k}{T}\right)+\cdots \right),
\label{eq:Rate}
\end{eqnarray}
where the dots denote higher order terms.
We plug \Eq{Rate} in \Eq{OmegaGWToday} and, in order to get an analytical result, we approximate $\gss(T)=\gsr(T)=\gsc(T)=\gs(\Tmax)$ in the region of temperatures above $\Tphi$. We thus obtain
\begin{eqnarray}
h^2_0\OmGW(\fg)=5.54\times 10^{-12}\left(\frac{\fg}{10^{10}\,{\rm Hz}}\right)^3\left(\frac{2}{\gs(\Tmax)}\right)^{\frac{5}{6}}\left(\frac{\Tmax/\Mp}{10^{-3}}\right) \nb(\ymax)\times\nonumber\\
\times\left(\lambda^2\psi^{(2,2)}\left(\ymax\right)+\frac{1}{3}\left(\frac{\Tmax}{\Mp}\right)^2\psi^{(0,4)}\left(y_{\rm max}\right)+\cdots \right),
\label{eq:OmegaGWToday_expansion}
\end{eqnarray}
where we have assumed $\Tmax\gg \Tphi$ and we have defined $\ymax\equiv\frac{2\pi\fg}{T_{\rm today}}\left(\frac{\gss(\Tmax)}{\gss(T_{\rm today})}\right)^{1/3}=0.14\left(\frac{\fg}{10^{10}\,{\rm Hz}}\right)\,\left(\frac{\gss(\Tmax)}{2}\right)^{1/3}$. 
Note that models which can describe the entire thermal history of the early universe have $\gs(\Tmax)>\gss(T_{\rm today})$. We work with a model that includes only one complex scalar field and therefore,  $\gs(\Tmax)<\gss(T_{\rm today})$. Nonetheless the features in the GW spectrum that we work out here will hold in general even with a more realistic model that can describe the thermal history of the universe consistently. We comment on this aspect further below.
The terms that are represented by the dots in \Eq{OmegaGWToday_expansion} include processes which encode quantum gravity effects, cf. \Fig{Ph_qg_effects}.
These effects arise at the order $\Otf$ and would appear as a term $\psi^{(2,4)}\lambda^2\left(\Tmax/\Mp\right)^2$ in the parantheses in the second line in \Eq{OmegaGWToday_expansion}.

The single graviton production processes have a matrix element squared of the order $\Ott$, i.e., it is proportional to $1/\Mp^2$. Since the GW production happens on a time scale that is comparable with $\Mp$, the single graviton production processes are proportional to $\Tmax/\Mp$ in the GW spectrum. A similar argument applies to the contribution from the graviton pair production. The matrix element squared is of the order $\Ozf$, i.e., proportional to $1/\Mp^4$, and hence the contribution to the GW spectrum is of the order $(\Tmax/\Mp)^3$. We refer to this production mechanism as GW freeze-in, since the GWs are produced from the thermal plasma throughout the expansion of the universe. Note that while the GW production mechanism is conceptually very similar to the production of dark matter from the thermal plasma, the final GW spectrum is very ultraviolet sensitive in the sense that it depends on the maximum temperature.

\begin{figure}[h]
	\centering
	\includegraphics[width=0.7\textwidth]{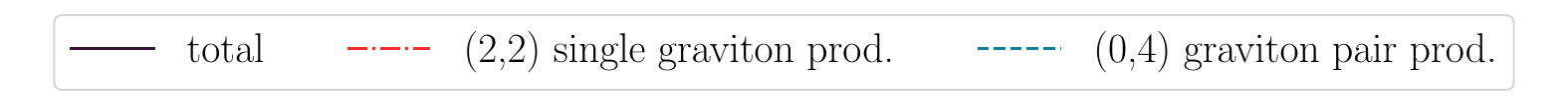}\\
	\includegraphics[width=0.49\textwidth]{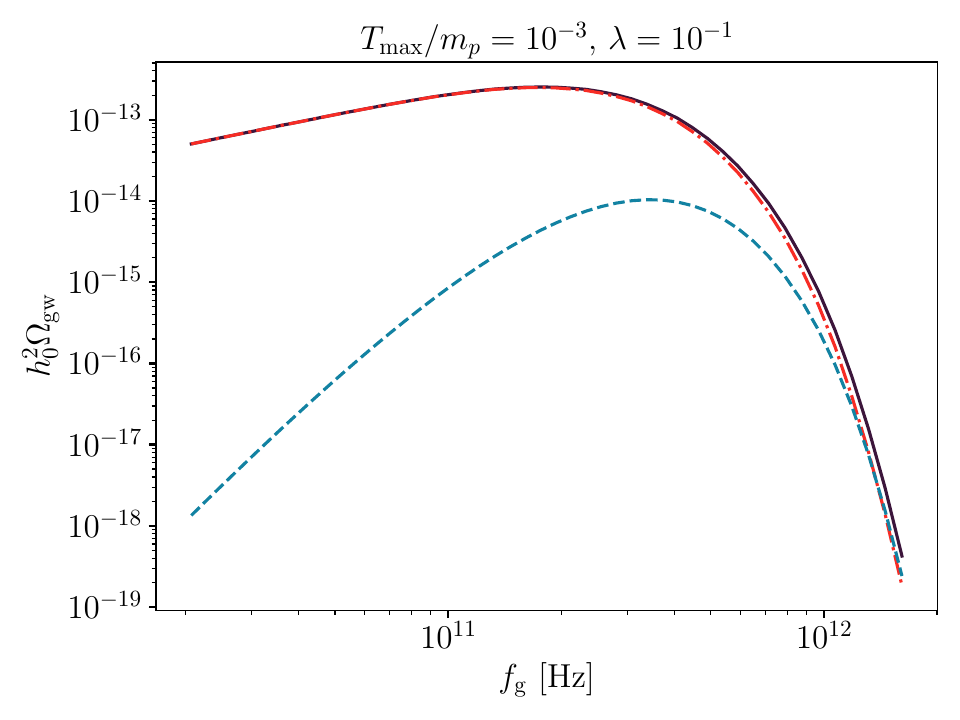}
	\includegraphics[width=0.49\textwidth]{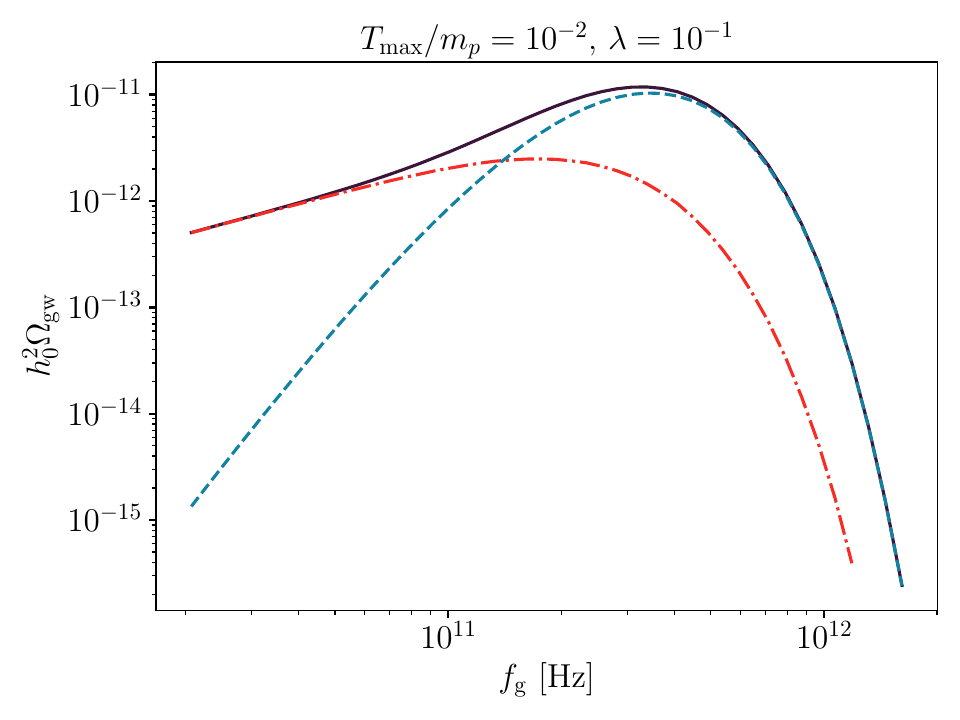}
	\caption{Gravitational wave spectrum with respect to the present day GW frequency. The single graviton production processes which are of order $\Ott=(2,2)$ are shown as a red dot-dashed line. The graviton pair production processes are shown as a dashed blue line and are of order $\Ozf=(0,4)$. The total GW spectrum is shown as a solid black line. 
	On the left we show a scenario where the maximum temperature is limited to $10^{-3}\, \Mp$. In the right figure we set the maximum temperature to $\Tmax=10^{-2}\, \Mp$. In this case the graviton pair production processes yield an even larger contribution compared to the single graviton production processes. The GW spectra are calculated for a complex scalar model with $\gs(\Tmax)=2$.}
	\label{fig:GW_spectrum}
\end{figure}

In \Fig{GW_spectrum} we plot the GW spectrum coming from  single graviton production processes (red dot-dashed lines) and from graviton pair production processes (blue dashed lines). The sum of both contributions, i.e., the total GW spectrum, is shown as a black solid line. The quartic coupling is always set to $\lambda=10^{-1}$.\footnote{In a more realistic model that can describe the entire history of our universe one would have to use renormalization group equations to run the parameters up to very high energy scales.}
When we evaluate the GW spectrum we also have to evaluate the $\psi$-functions. Since these functions have only been calculated reliably for arguments that are larger than $\sqrt{\lambda}=0.31$, cf. \Sec{Matrix_elements}, we only show the GW spectrum in the corresponding frequency regime.
Note that the GW spectrum from the single graviton production mimics  to a good approximation a black body spectrum, since the function $\psi^{(2,2)}$ is very flat in the regime $T/\Mp>\sqrt{\lambda}$. The graviton pair production contribution has a significantly different shape since the function $\psi^{(0,4)}$ is not constant in the frequency interval of interest. 

\Fig{GW_spectrum} (left) shows a scenario where the maximum temperature is set to $\Tmax/\Mp=10^{-3}$. 
In this case the single graviton production contribution  dominates over the graviton pair production contribution. 
The total spectrum has therefore mostly the form of the single graviton production spectrum.

\Fig{GW_spectrum} (right) shows the GW spectrum for a slightly larger maximum temperature, $\Tmax/\Mp=10^{-2}$, which requires a non inflationary scenario. Note that the chosen maximum temperature is consistent with the equilibrium condition for the scalar fields, i.e. $\lambda^2\gtrsim T/\Mp$. In the case at hand both contributions are parametrically equally important and around the peak frequency the graviton pair production contribution
is even substantially larger than the single graviton production contribution. This can be seen explicitly from \Eq{OmegaGWToday_expansion} by comparing the two terms in the second line. The first term corresponds to single graviton production processes, while the second one describes the contribution due to graviton pair production. Comparing both terms we find as an order of magnitude estimate, that the contribution from graviton pair production processes are equally important or even larger than the contribution from the single graviton production processes if $10 \,\Tmax/\Mp \gtrsim \lambda$.
Therefore, the relevance of the graviton pair production processes depends crucially
on the size of the coupling and the maximum temperature.
The GW spectra associated with single graviton and graviton pair production processes peak at slightly different frequencies and have a distinct functional form. As a result, the total spectrum takes a very characteristic form that is substantially different from the single graviton production spectrum, i.e., an approximate black body spectrum.

In the following we derive analytic expressions for the peak frequencies of the single graviton and graviton pair production GW spectra.
For the single graviton contribution we find from \Eq{OmegaGWToday_expansion} that the GW spectrum peaks at: $2 \times 10^{11}\,{\rm Hz}\, \left(2/\gss(\Tmax)\right)^{1/3}$, where we have assumed that $\psi^{(2,2)}=\,$const. which is a good approximation in the frequency interval of interest. The peak frequency of the graviton pair production curve lies at a slightly higher frequency: $3.5 \times 10^{11}\,{\rm Hz} \,\left(2/\gss(\Tmax)\right)^{1/3}$, where we have used the asymptotic form $\psi^{(0,4)}=32/(15 \pi) (k/T)^2$ that was derived in \App{integrate}.

The higher order terms which are depicted by the dots in \Eq{OmegaGWToday_expansion} are suppressed further by powers of $\Tmax/\Mp$ and $\lambda$. 
For values of $\Tmax/\Mp$ and $\lambda$ that are relatively close to unity one has to check in detail that the higher order $\psi$ functions are not larger than the leading order $\psi$ functions such that the suppression from the additional powers of $\Tmax/\Mp$ and $\lambda$ is not spoiled. The values of $\Tmax$ and $\lambda$, that we consider in \Fig{GW_spectrum} are much smaller than unity and therefore we do not expect such a scenario to happen. For example the $\psi^{(0,6)}$ function would have to be four orders of magnitude larger than the $\psi^{(0,4)}$ function for $\Tmax/\Mp=10^{-2}$. The higher order $\psi$ functions will be either phase space suppressed or have the same phase space as the leading order processes. 
In both cases we do not expect an enhancement of the higher order $\psi$ functions by orders of magnitude.

The maximum of the CGMB spectrum is bounded from above, $h_0^2\OmGW\lesssim 10^{-6}$~\cite{Ringwald:2020ist}, due to constraints on the additionally allowed amount of dark radiation. Both scenarios that are shown in \Fig{GW_spectrum} do not saturate this bound and are therefore not excluded. 
In the complex scalar model the dark radiation bound is saturated for $\Tmax/\Mp \simeq 0.5$. Note that in this case the main contribution to the GW spectrum is coming from graviton pair production processes, which illustrates the importance to include these processes at high temperatures.
The SM predictions for the GW production from the thermal plasma includes currently only single graviton production processes, cf. Refs.~\cite{Ghiglieri:2015nfa,Ghiglieri:2020mhm,Ringwald:2020ist}, and yields $h_0^2\OmGW=5\times 10^{-7}$ for $\Tmax=\Mp$. Therefore, adding graviton pair production processes to the SM calculation can already lead to a violation of the dark radiation bound for $\Tmax<\Mp$, because the current prediction which includes only single graviton production processes, is already very close to the dark radiation bound. In conclusion, this might then be used to constrain the maximum temperature of the universe. We plan to work out the details in a follow-up study where we will also address the aspect of thermal equilibrium at ultra high temperatures in the SM and beyond the SM theories.

\section{Conclusions and Outlook}\label{sec:conclusion}

The thermal plasma in the early universe produced a guaranteed stochastic GW background through thermal fluctuations. 
At each time the emitted GW spectrum peaks at the respective temperature. Due to the temperature-redshift relation, the peak frequencies of the GW spectra are all redshifted to the same frequency today and therefore add up.
Conceptually, the GW production from the thermal plasma has many similarities with the so-called dark matter freeze-in production from the thermal plasma. The GWs are produced out of equilibrium and their distribution function is small at all times, $\fh\ll 1$. Furthermore, the $\fh$ distribution function evolves much slower than the Hubble rate. 
We therefore dubbed the GW production from the thermal plasma \textit{GW freeze-in production}.
The GW freeze-in scenario is ultraviolet dominated in the sense that it depends on the maximum temperature of the universe, as expected from a non-renormalizable coupling.

In this paper we use a Boltzmann-like formalism to study the microscopic particle collision processes that contribute to the CGMB spectrum. We have done all calculations in a model  with a complex scalar field and quartic self-interaction. Our basic assumption is that after the hot Big Bang a plasma of scalars with temperature $\Tmax$  is present and this plasma produced the CGMB spectrum.
First, we considered the contribution of single graviton production processes to the CGMB spectrum. 
In a scalar theory with quartic interaction, single graviton production processes are $2\leftrightarrow 3$ processes, which have not been calculated before.
Our calculation is motivated by the fact that a quartic coupling exists in the Higgs sector of the SM and in many BSM theories. 
The second class of processes that we investigate are graviton pair production processes. These are $2\to 2$ processes and have not been considered before in the context of GWs from the thermal plasma.
We show that their contribution to the CGMB spectrum can be larger than the contribution from the single graviton production processes.
As an order of magnitude estimate, graviton pair production processes dominate the GW spectrum if $10 \,\Tmax/\Mp \gtrsim \lambda$.
Note however that the maximum temperature is also bounded from above by an equilibrium requirement for the scalar particles: $\lambda^2\gtrsim \Tmax/\Mp$, which has to be seen as a parametric estimate.
Therefore, the degree to which graviton pair production processes contribute significantly to the CGMB spectrum depends on the values of the coupling coefficient and the maximum temperature.
As an example we show the two different scenarios in \Fig{GW_spectrum}. On the  left single graviton production processes dominate ($\Tmax/\Mp=10^{-3}$ and $\lambda=10^{-1}$). When increasing $\Tmax$ by one order of magnitude (\Fig{GW_spectrum}, right) graviton pair production processes yield a significant contribution to the total GW spectrum.

The single graviton and graviton pair production processes are the lowest order contributions which can easily be incorporated into our Boltzmann-like formalism. We have also discussed the first steps and problems that would arise if one would add real and virtual quantum gravity corrections to the presented results. While finite real corrections can be incorporated in our formalism, virtual corrections depend on the distribution functions themselves and this complicates their inclusion into the Boltzmann-like approach that we use here. 
A possible future direction is to  derive a quantum Boltzmann equation from the Wigner function by performing a systematic $\hbar$ expansion. This would allow one to explicitly identify the quantum corrections.

The results that we have worked out for a scalar model are qualitatively also valid for more general theories. 
In this case the coupling coefficient $\lambda$ would have to be replaced with the heat bath couplings in the more general theory, which we generically refer to as $g$. 
Then the contribution from graviton pair production processes to the GW spectrum
dominates over the single graviton contribution
if $X\times \Tmax/\Mp\gtrsim g$, where $X$ is a model dependent constant. A value $X<10$ in the SM would indicate that graviton pair production dominates at higher temperatures compared to our scalar model. In a follow-up study we will answer this point with a full SM and BSM calculation. 
Confronting the graviton pair production calculation in the SM and BSM theories with existing dark radiation constrains can therefore already lead to constraints on $\Tmax$. Constraints on $\Tmax$ can be used to test different models of our universe, cf. Ref.~\cite{Ringwald:2020ist}. For example non-standard inflationary cosmological models would be required if a $\Tmax \gtrsim 10^{-3}\,\Mp$ would be inferred from the GW spectrum. Furthermore, the CGMB can be used to constrain non-standard cosmological histories, cf. Ref.~\cite{Muia:2023wru}. The authors of Ref.~\cite{Muia:2023wru} have only considered single graviton production processes. It would be interesting to add graviton pair production processes to their calculation since it could lead to stronger constraints on non-standard cosmological histories.

A detection of the CGMB with Earth-based detectors will be challenging, cf. Ref.~\cite{Berlin:2021txa}, however there exist detector proposals with sensitivities comparable to the dark radiation bound, cf. Ref.~\cite{Aggarwal:2020olq}. Future experimental work will have to show if the proposed detectors can be realized and if their foreseen sensitivity can even be improved. 
Our results motivate further work on high-frequency GW detection since a detection of the CGMB in the future would pave the way to probe our understanding of particle physics and cosmology at ultra high energies.

\section*{Acknowledgments}
We thank Alex Dima, Yoni Kahn, Mikko Laine, Kaloian Lozanov, Daniel Meuser, Patrick Peter and Andreas Ringwald for useful discussions. We also thank Nick Abboud, Rachel Nguyen and Michael Wentzel for linguistic corrections.
The work of JSE is supported in part by DOE grant DE-SC0015655. 
JG acknowledges support by a PULSAR grant from the R\'egion Pays de la Loire.
JSE would like to express special thanks to the Mainz Institute for Theoretical Physics (MITP) of the Cluster of Excellence PRISMA+ (Project ID 39083149), for its hospitality and support.

\appendix

\section{Details on the  evaluation of the phase space integrals}
\label{app:integrate}
In this appendix we provide some extra details on the
phase space integrals of Sec.~\ref{sec:Matrix_elements}.
Let us start from Eqs.~\eqref{kin_thy23_expl} and \eqref{kin_thy32expl}.
In order to carry out the integrations numerically, we can 
rewrite the phase space as 
\begin{align}
    \int \! {\rm d}\Omega^{ }_{2\to3}=-\frac{1}{16(2\pi)^7}
    \int_0^\infty dp'_1\,p'_1\int_0^\infty dp'_2\,p'_2
    \int_{-1}^1 d c_{p'_1}
    \int_{-1}^1 d c_{p'_2}
    \int_{-1}^1 d c_{p_1}
    \int_0^{2\pi}d \phi_1 \int_0^{2\pi}d \phi_2\, \frac{p_1^2\theta(p_1)
    \theta(p'_1+p'_2-p_1-k)}{P_1\cdot P_2},
    \label{phase_expl}
\end{align}
where $k$ is chosen to point in the $z$ direction and
the $c$ variables are the cosines of the angles between $k$ and the respective momenta: $c_p\equiv\cos\theta_{k,p}$.
The $\phi$'s are two azimuthal angles, where the third one was integrated out. $p_1$ is fixed to
\begin{align}
    p_1=&\big[P'_1\cdot P'_2-P'_1\cdot K-P'_2\cdot K\big]\big[k(1-c_{p_1})
    -p'_1(1-(c_{p'_1}c_{p_1}+\sqrt{(1{-}c_{p'_1}^2)(1{-}c_{p_1}^2)}\cos(\phi_1{+}\phi_2)))\nonumber \\
    &  -p'_2(1-(c_{p'_2}c_{p_1}+\sqrt{(1{-}c_{p'_2}^2)(1{-}c_{p_1}^2)}\cos(\phi_2)))\big]
\end{align}
and $P'_i\cdot K=p'_ik(c_{p'_i}-1)$, $P'_1\cdot P'_2=p'_1p'_2(c_{p'_1}c_{p'_2}+\sqrt{(1{-}c_{p'_1}^2)(1{-}c_{p'_2}^2)}\cos(\phi_1)-1)$. Similarly, $P'_i \cdot P_1=
p'_ip_1(c_{p'_i}c_{p_1}+\sqrt{(1{-}c_{p'_i}^2)(1{-}c_{p_1}^2)}\cos(\delta_{i1}\phi_1{+}\phi_2)-1)$.
The other inner products, including $P_1\cdot P_2$, follow from $P_2=P'_1+P'_2-K-P_1$.
The $3\to 2$ analog of Eq.~\eqref{phase_expl} follows from simple crossings.
These seven-dimensional integrals are carried out numerically
using the Monte Carlo algorithm \textsc{Vegas+}~\cite{Lepage:2020tgj}.
The results are shown in \Fig{fig_phi_functions}.

Let us now consider the $\Ozf$ contribution. \Eq{kin_thy22_expl}
can be evaluated using the standard ``s-channel'' parametrisation of Ref.~\cite{Baym:1990uj,Besak:2012qm}.
We can arrange the phase space integral as
\begin{equation}
  \label{schannel}
  \int \! {\rm d}\Omega^{ }_{2\to2} =\frac{1}{(4\pi)^3k}\int_k^\infty d q^0 \int_{\vert 2k-q^0\vert}^{q^0} dq \int_{q_-}^{q_+}dp'_2\int_0^{2\pi}\frac{d\phi_{p'_2k}}{2\pi} 
\end{equation}
where we defined $q_\pm\equiv(q^0\pm q)/2$ and we chose $q^0,q$ such that
$p'_1=q^0-p'_2$ and $\mathbf{p}'_1+\mathbf{p}'_2=\mathbf{q}$. $\phi_{p'_2k}$
is the azimuthal angle between the $\mathbf{p}'_2,\mathbf{q}$ and  $\mathbf{k},\mathbf{q}$
planes.
This corresponds to
\begin{equation}
    s=q_0^2-q^2,\quad t=-\frac{s}{2q^2}\bigg[(2k-q^0)(q^0-2p'_2)+q^2-\cos(\phi_{p'_2k})
    \sqrt{(q^2-(q^0-2k)^2)(q^2-(q^0-2p'_2)^2)}\bigg].
\end{equation}
We can perform the angular average $\langle\ldots\rangle_{\phi_{p'_2k}}\equiv \int_0^{2\pi}\frac{d\phi_{p'_2k}}{2\pi} \ldots$ to get rid of odd powers of the cosine and find
\begin{equation}
  \label{kin_thy22_expl3}
  \dfh^{(0,4)}=\frac{\kappa^4}{32k}
   \frac{1}{(4\pi)^3k}\int_k^\infty d q^0 \int_{\vert 2k-q^0\vert}^{q^0} dq \int_{q_-}^{q_+}dp'_2 \bigg\langle t^2\bigg[\frac{t^2}{s^2}+2\frac{t}{s}+1\bigg]\bigg\rangle_{\phi_{p'_2k}}
  \nB(q^0)\,[1+\nB(p'_2)+\nB(q^0-p'_2)]
  \;.
\end{equation}
We have used the identity $\nB(p'_2)\nB(q^0-p'_2)=\nB(q^0)\,[1+\nB(p'_2)
+\nB(q^0-p'_2)]$ which is useful for treating the $p'_2$ integration analytically.\footnote{The matrix element squared depends on $p'_2$ as function of $q^0-2p'_2$. It thus has a reflection symmetry around $p'_2=q^0/2$,
which is the midpoint of the $p'_2$ integration range. We can then reflect the $\nB(q^0-p'_2)$
into an $\nB(p'_2)$  leaving the matrix element unchanged, further simplifying the integration.}
Carrying out the integration we find:
\begin{align}
 \dfh^{(0,4)}=&\frac{\kappa^4}{4(8\pi)^3k^2}\int_k^\infty d q^0 \int_{\vert 2k-q^0\vert}^{q^0} dq\, \nB(q^0) \,
(q_0^2-q^2)^2
\bigg\{-\frac{11 q^4-30 q^2 (2k-q^0)^2+15 (2k-q^0)^4}{120 q^3}\nonumber\\
 &+T\frac{(q^2-(q^0-2k)^2)^2}{8q^4}   \ln\frac{
  e^{\frac{q_+}{T}}-1}{e^{\frac{q_-}{T}}-1}\nonumber\\
&-T^2\frac{ (q^2-(q^0-2k)^2)(q^2-5(q^0-2k)^2) }{2q^5}
   \left(\text{Li}_2\left(e^{-\frac{q_+}{T}}\right)+\text{Li}_2\left(
   e^{-\frac{q_-}{T}}\right)\right)\nonumber \\
   &-T^3\frac{ 45 (2 k - q^0)^4 + q^2 (5 q^2 - 42 (q^0 - 2 k)^2)
   }{2q^6}
   \left(\text{Li}_3\left(e^{-\frac{q_+}{T}}\right)-\text{Li}_3\left(
   e^{-\frac{q_-}{T}}\right)\right)\nonumber \\
    & -T^4\frac{9 q^4-90 q^2 (2k-q^0)^2+105 (2k-q^0)^4}{q^7}
   \left[\text{Li}_4\left(e^{-\frac{q_+}{T}}\right)+\text{Li}_4\left(
   e^{-\frac{q_-}{T}}\right)+\frac{2T}{q}\left(\text{Li}_5\left(e^{-\frac{q_+}{T}}\right)-\text{Li}_5\left(
   e^{-\frac{q_-}{T}}\right)\right)\right]\bigg\}.
   \label{expl2d}
\end{align}
Note that the result of the integration is positive, though this might not appear
obvious from this expression.
For large momenta, $k\gg T$, Eq.~\eqref{expl2d} asymptotes to 
\begin{equation}
     \dfh^{(0,4)}\big\vert_{k\gg T}=\frac{\kappa^4T^4}{15(4\pi)^3}e^{-k/T}\bigg(k+\mathcal{O}\left(\frac{1}{k^2}\right)\bigg)\,.
     \label{eq:asymptotic}
\end{equation}
This result can be extracted by noting that in this asymptotic regime,
$\nB(q^0)\approx e^{-q^0/T}$. This sharp exponential cutoff 
ensures that only the $q^0\approx k$ and $2k-q^0<q<q^0$ ($q\approx k$) 
ranges dominate the integral. Expanding the integrand for $q^0-k\ll k$
and $q-k\ll k$ and then performing the integral we recover \Eq{asymptotic}. 
We note that 
the form given in \Eq{asymptotic}, while valid for $k\gg T$, approximates
the numerical results shown in \Fig{fig_phi_functions} at better than
30\% accuracy for $k>T$.

\bibliography{bibliography}

\end{document}